\documentclass[preprint, 12pt]{elsarticle}

\usepackage[margin = 2cm]{geometry}
\usepackage{graphicx, subfig}
\usepackage{amssymb, amsthm, amsmath}

\DeclareMathOperator{\E}{\mathbb{E}}                 
\DeclareMathOperator{\Var}{Var}                      
\DeclareMathOperator{\Cov}{Cov}                      
\DeclareMathOperator{\Prob}{\mathbb{P}}              
\DeclareMathOperator{\Rg}{Rg}                        

\newcommand{\N}{\mathbb{N}}                          
\newcommand{\R}{\mathbb{R}}                          
\newcommand{\Z}{\mathbb{Z}}                          
\newcommand{\T}{\mathcal{T}}                         
\newcommand{\set}[1]{\lbrace#1\rbrace}               
\newcommand{\normtwo}[1]{\lVert#1\rVert_2}             
\newcommand{\normRd}[1]{\lVert#1\rVert_{\R^d}}       
\newcommand{\hilnorm}[1]{\lVert#1\rVert_H}           
\newcommand{\supnorm}[1]{\lVert#1\rVert_\mathcal{L}} 
\newcommand{\hsnorm}[1]{\lVert#1\rVert_{\mathcal{K}_2}} 
\newcommand{\innH}[1]{<#1>_H}                        
\newcommand{\Int}{\int_0^1}

\newcommand{\hata}{\widehat{a}_{v, n}}
\newcommand{\hatf}{\widehat{f}_{n}}
\newcommand{\hatg}{\widehat{g}_{v, n}}

\newcommand{\arh}{{\sc arh}} 
\newcommand{\carh}{{\sc carh}} 

\newtheorem{thm}{Theorem}[section]
\newtheorem{cor}[thm]{Corollary}
\newtheorem{lem}[thm]{Lemma}
\newtheorem{prop}[thm]{Proposition}
\theoremstyle{definition} 
\newtheorem{assumption}[thm]{Assumptions}

\journal{Journal of Multivariate Analysis}

\begin{document}

\begin{frontmatter}


\title{Conditional Autoregressive Hilbertian processes}
\author{Jairo Cugliari}
\ead{Jairo.Cugliari@inria.fr}
\address{INRIA Research Team Select, 
         Universit\'e Paris Sud B\^at. 425, 91405 Orsay Cedex, France}

\begin{abstract}
When considering the problem of forecasting a continuous-time stochastic process over an entire time-interval in terms of its recent past, the notion of Autoregressive Hilbert space processes (\arh) arises. This model can be seen as a generalization of the classical autoregressive processes to Hilbert space valued random variables. Its estimation presents several challenges that were addressed by many authors in recent years.

In this paper, we propose an extension based on this model by introducing a conditioning process on the \arh.  In this way, we are aiming a double objective. First, the intrinsic linearity of \arh~is overwhelm. Second, we allow the introduction of exogenous covariates on this function-valued time series model.

We begin defining a new kind of processes that we call Conditional \arh. We then propose estimators for the infinite dimensional parameters associated to such processes. Using two classes of predictors defined within the \arh~framework, we extend  these to  our case. Consistency results are provided as well as a real data application related to electricity load forecasting. 
\end{abstract}

\begin{keyword}
Functional Data \sep Nonparametric \sep Forecasting \sep Exogenous covariate
\MSC 62G08 \sep 62M10
\end{keyword}

\end{frontmatter}



   \section{Introduction}\label{sec:intro_carh}

We consider a function-valued process $Z = (Z_k, k\in\Z)$ where for each $k$, $Z_k$ is a random element taking his values in some functional space $F$. A popular choice is to set $F = H$ a real separable Hilbert space because of the rich geometric properties of Hilbert spaces. As for classical time series, an important task is the problem of obtaining some information about the future value $Z_{n+1}$ from the observed discrete sequence $Z_1,\ldots,Z_n$. Then, the best predictor (in the quadratic mean loss function sense) of the future observation $Z_{n+1}$ is its conditional expectation given the past
\begin{equation}\label{eq:pred3}
  \tilde{Z}_{n+1} = \E ( Z_{n+1} | Z_n,\ldots, Z_1) ,
\end{equation}
which may depend on the unknown distribution of $Z$.

One important case arises when one assumes that $Z$ is a strictly stationary zero-mean Autoregressive Hilbertian process of order 1  \arh(1), introduced by \citet{bosq1991modelization} and defined by
\begin{equation}\label{eq:arh}
    Z_{k + 1} = \rho Z_{k} + \epsilon_k, \qquad k\in\Z,
\end{equation}
with $\rho$ a bounded linear operator over $H$ and $\epsilon = (\epsilon_k, k\in\Z)$ a strong $H$-valued white noise. For this process, the best predictor of $Z_{n+1}$ given the past observations is $\tilde{Z}_{n+1} = \rho Z_n$. Notice that $\rho$ is usually unknown. Two forecasting strategies can be followed here. The first one is to first estimate $\rho$ and then apply it to the last observation $Z_n$ to obtain a prediction of $Z_{n+1}$ (see \citet{bosq1991modelization, besse1996approximation, pumo1998prediction}). Alternatively, one may directly predict $\tilde{Z}_{n+1}$ by estimating the relevant elements of the range of $\rho^{*}$ (see  \citet{antoniadis2003wavelet}). Adopting this last strategy and using some wavelet decomposition the later authors obtain considerable better prediction results. The choice of a wavelet basis is guided by the good approximation properties they have to represent quite irregular trajectories of $Z$. \citet{kargin2008curve} also use the second strategy but propose to use a data-dependent basis adapted to the prediction task. 

While \arh~processes are a natural generalization of the well known autoregressive processes in Euclidean spaces, the infinite dimension of the space $H$ produces new challenges for their estimation and prediction (see \citet{mas2009linear} for a recent review on this topic). A second issue is the study of some of the extensions developed on the scalar case to the Hilbertian framework, like for instance higher order \arh~processes studied in \citet{pumo1992prediction}. 

We are interested in another extension taking into account some exogenous information modeled by the influence of covariates in the model given by equation \eqref{eq:arh}. We may cite \citet{mas2007arhd} that uses  the derivative of $Z_k$ as a covariate, or \citet{damon2002inclusion} that introduces a function-valued covariate also following an \arh~process. In both these works, the covariates are introduced as additive terms in the equation \eqref{eq:arh}.

Alternatively, one may introduce exogenous information through the linear operator $\rho$. Like in the scalar case, one may consider a more general case where the parameter $\rho$ depends on some covariate. For such cases, the exogenous information may be incorporated in a non-additive manner. \citet{guillas2002doubly} propose to model $Z$ by a doubly stochastic Hilbert process defined by 
\begin{equation}\label{eq:carh_gui}
   Z_k = \rho_{V_k} (Z_{k-1}) + \epsilon_k, \qquad k\in\Z,
\end{equation}
where $V = (V_k, k\in\Z)$ is a sequence of independent identically distributed Bernoulli variables. The intuition behind the model is that there exists two regimes expressed through two different operators, $\rho_0$ and $\rho_1$. At each instant $k$, one of the regimes is randomly chosen as the result of the drawn of the associate Bernoulli variable $V_k$. The resulting process admits to have one of the regimes to be explosive if it is not visited too often. In such a case, equation \eqref{eq:carh_gui} has a unique stationary solution. 

In this paper, we introduce the \textit{Conditional Autoregressive Hilbertian process} (\carh), constructed such that conditionally on an exogenous covariate $V$, the process $Z$ follows an \arh~process. While \carh~definition is similar to equation \eqref{eq:carh_gui}, it  differs mainly in two ways. The first one is related to the nature of the process $V$ which we assume to be a multivariate random process with some continuous distribution. Second, we propose predictors that will accomplish the prediction task using the exogenous information. Indeed, the exogenous information of the actual regime is used to found similar local situations on the observed past. 

The paper is structured as follows. In Section \ref{sec:carh} we introduce the main definitions and we present the model. Linear operators on Hilbert spaces are intensively used through out the article. On \ref{sec:lops} we recall some important facts on this topic that we use on the article. We also propose estimators for the unknown parameters as well as two classes of predictors. The main results about the convergence of the estimators and predictors are shown in Section \ref{sec:main} postponing the proofs until the \ref{sec:proofs}. Finally, Section \ref{sec:empirical} contains a real data application of  \carh~processes which illustrates empirically the performance of the predictors.

 \section{Conditional \arh~process: \carh.}\label{sec:carh}

After some notations, we define the \carh~process and we propose estimators for the associated parameters operators. Then, we follow  prediction strategies similar to those adopted in previous studies  for \arh~processes,   to obtain classes of predictors for the \carh~process.

\subsection{Preliminaries}

All  variables are defined on the same probability space $(\Omega, \mathcal{F}, \Prob)$. We consider a sequence $Z = (Z_k, k\in\Z)$ of Hilbert space valued random variables, i.e. each random variable $Z_k$ is a measurable map from the probability space in an real separable Hilbert space $H$ endowed with its Borel $\sigma$-field, $\mathcal{H}$. The space $H$ is equipped with the scalar product $\innH{., .}$ and the induced norm $\|.\|_H$). We also consider a sequence of real $\R^d-$valued random variables $V = (V_k, k\in\Z)$. Both sequences $Z$ and $V$ are assumed to be stationary. We will focus on the behaviour of $Z$ conditionally on $V$. We will further assume that $Z$ is strongly integrable.

The conditional expectation is characterized by the conditional distribution of $Z$ given $V$, i.e. by the conditional probability $\Prob_{Z|V}$ on $\mathcal{H}$. In order to ensure that this conditional probability is properly defined as a measure (in the sense that it represents a regular version of the conditional probability), it is assumed that a transition probability exists that associates to each $v\in\R^d$ a probability measure $\Prob^v$ on $(H, \mathcal{H})$ such that 
  \[ \Prob_{Z|V}^v (A) = \Prob^v(A), \qquad \text{for every } A \in \mathcal{H}, v\in\R^d.  \]
We call $\Prob^v$ the sampling measure and denote $\E^v$ the induced expectation. We restrict our attention to functions defined over a real compact interval $\T$ and we assume hereafter $T$ to be $[0, 1]$ without loss of generality. More precisely, we set $H$ to be the subspace of continuous functions on the space of classes of 4-th order $\Prob-$integrable functions. 

\subsection{The model}

A sequence $(Z, V) = \{(Z_k,V_k), k\in\Z \}$ of $H \times \R^d$-valued random variables is a Conditional Autoregressive Hilbertian process (\carh) of order 1 if it is stationary and such that, for each $k$
\begin{equation}\label{eq:carh}
  Z_k = a + \rho_{V_k} (Z_{k - 1} - a) + \epsilon_k ,
\end{equation}
where the conditional mean function $a_v = \E^v [Z_0 | V], v\in\R^d,$ is the conditional expectation (on $V$) of the process, $\epsilon = (\epsilon_k, k\in\Z)$ is an $H-$valued white noise and $(\rho_{V_k}, k\in\Z)$ is a sequence of random operators such that, conditionally on $V,$ $\rho_{V}$ is a linear compact operator on $H$ (see \ref{sec:lops}). Additionally, $V$ and $\epsilon$ are independent process. Using the following assumptions we prove the existence and uniqueness of the \carh~processes~(see \ref{sec:proofs} for the proof).

\begin{assumption}\label{ass:existence}
Assume that:
\begin{enumerate}
\item There exists a map $v \mapsto P^v$ that assigns a probability measure on $(H, \mathcal{H})$ to each value $v$ in the support of $V$. 
\item $\sup_n \|\rho_{V_n}\|_\mathcal{L} = M_\rho < 1$ .
\end{enumerate}
\end{assumption}
\begin{thm}\label{teo:existence}
Under Assumptions \ref{ass:existence}, equation \eqref{eq:carh} defines a \carh~process with an unique stationary solution given by 
\[ Z_k = a + \sum_{j=0}^\infty 
         \left(\prod_{p = 0}^{j - 1} \rho_{V_{k - p}} \right) (\epsilon_{k - j}) ,\]
with the convention $\prod_{p = 0}^{j - 1} \rho_{V_{k - p}} = \text{Id}$ (the identity operator) for $j=0$. 
\end{thm}

The first condition in Assumption \ref{ass:existence} has already been discussed. The second one ensures the contraction of the conditional autoregressive operator through the supremum norm of linear operators (see \ref{sec:lops}).

\subsection{Associated operators}

Hereafter we make the additional assumption that $\E^v [ \hilnorm{Z}^4 |V] < \infty$. Let us note $H^*$ the topological dual of $H$, i.e.~the space of bounded linear functionals on $H$. We introduce two linear operators mapping from $H^*$ to $H$ associated to the \carh~process. Thanks to the Riesz representation, $H^*$ the topological dual of $H$ can be identified with $H$, and the operators may be defined as follows:
\begin{align*}
 z \in H \mapsto \Gamma_v z &= 
      \E^v [((Z_0 - a) \otimes (Z_0 - a))(z) | V ] \qquad\text{and} \\
 z \in H \mapsto \Delta_v z &= 
      \E^v [((Z_0 -a ) \otimes (Z_1 - a))(z) | V ] ,
\end{align*}
that we call \textit{conditional (on $V$) covariance} and \textit{cross covariance operators} respectively. We have used the tensor product notation $(u \otimes v)(z) = \innH{u, z} v$ for $u, v, z \in H$. 

For each $v\in\R^d$, both $\Gamma_v$ and $\Delta_v$ are trace-class and hence Hilbert-Schmidt. In addition, $\Gamma_v$ is positive definite and self adjoint. Then, we may write down the spectral decomposition of $\Gamma_v$ as
  \[ \Gamma_v = \sum_{j\in\N} \lambda_{v, j} (e_{v, j} \otimes e_{v, h} ) \]
where $(\lambda_{v, j}, e_{v, j})_{j\in\N}$ are the eigen-elements of $\Gamma_v$. The eigenvalues may be arranged to form a non-negative decreasing sequence of numbers tending towards zero. 

As a direct consequence of the choice  made for $H$, the operators have associated kernels $\gamma_v$ and $\delta_v$ defined over $L_2 ([0, 1]^2)$ such that
\begin{align*}
 \Gamma_v (z )(t) & = \Int \gamma_v(s, t) z(s) ds, \\
 \Delta_v (z) (t) & = \Int \delta_v(s, t) z(s) ds, \qquad 
  t \in [0, 1], v\in\R^d, z\in H,
\end{align*}
with $\gamma_v(., .)$ a continuous, symmetric and positive kernel and $\delta_v( ., .)$ a continuous kernel. The kernels turn to be the \textit{conditional covariance function} $\gamma_v( s, t) = \E^v [(Z_0(s) - a(s))(Z_0(t) - a(t))|V]$, and the \textit{one-step-ahead conditional cross covariance function} $\delta_v(s, t) = \E^v [(Z_0(s)-a(s))(Z_1(t)-a(t))|V]$, $(s,t) \in [0, 1]^2, v\in\R^d$. 

A Yule-Walker like relation links the operators $\Delta_v, \Gamma_v$ and $\rho_v$. For each $v\in\R^d$ we have
  \begin{equation}\label{eq:covrho}
    \Delta_v = \rho_v \Gamma_v . 
  \end{equation}
Using the property of the adjoint and the symmetry of $\Gamma_v$, we obtain from \eqref{eq:covrho} the following key relation for the estimation of $\rho_v$ (see Section 2.5),
\begin{equation}\label{eq:covrhoADJ}
   \Delta^*_v = \Gamma_v \rho^*_v .
\end{equation}

\subsection{Estimation of $a, \Gamma_v, \Delta_v$.}

The parameters can be estimated from data. We call $\{(Z_1, V_1), \ldots, (Z_n, V_n)\}$ the observed data supposed to come from a \carh~process. We use nonparametric Nadaraya-Watson like estimators to estimate the infinite-dimensional parameters $a_v, \Gamma_v$ and $\Delta_v$. This is a popular choice when the the parameters are defined through conditional expectations.

\subsubsection{Estimation of $a_v$.}

We estimate the conditional mean function of the process $a_v(t) = \E^v[Z_0 (t)| V]$ for all $t\in[0, 1]$ using the observations $\lbrace (Z_1, V_1), \ldots, (Z_n, V_n)\rbrace$. In order to properly define the framework, let us introduce some quantities. For some fixed $t \in [0, 1]$, set $Y = Z_0(t)$ and $Y_i = Z_i(t), i = 1, \ldots, n$. Let us assume that the distribution of $V$ admits a density $f$ with respect to the Lebesgue measure. We define for $v\in\R^d$
    \[ g_v(t) =  \E^v [Z_0(t) f(V) | V], \]
and provided that $f(v)>0$ we rewrite the parameter as the regression of $Y$ against $V$, 
\begin{align*} 
  a_v(t)  & =  \E^v [Y |V]          \\ 
          & =  \frac{g_v(t)}{f(v)} .  
\end{align*}
When $f(v) = 0$ we set $a_v(t) = \E [Y]$. 

The introduced quantities can be estimated by Nadaraya-Watson kernel based estimators. In our case, we use the following estimators for $f$ and $g_v$ respectively, 
\begin{align} \label{eq:hatf}
 \widehat f_n(v)   & = \frac{1}{n h_a^d} 
   \sum_{i=1}^n K (h_a^{-1} (V_i - v)) \qquad \text{and} \\
 \widehat g_{v, n} (t) & = \frac{1}{n h_a^d} 
   \sum_{i=1}^n K (h_a^{-1} (V_i - v)) Y_i , \label{eq:hatg}
\end{align}
where $K:\R^d \mapsto \R$ is a unitary square-integrable $d-$dimensional kernel and the bandwidth $h_a = (h_{a, n})_{n\in\N}$ is a decreasing sequence of positive numbers tending to 0 called the bandwidth. The estimator of $a_v(t)$ is then given by 
\begin{align*}
 \widehat a_{v, n} (t) & = \frac{\widehat g_{v, n}(t)}{\widehat f_n(v) },
\end{align*}
which can be written as $\widehat a_{v, n} = \sum_{i=1}^n w_{n, i}(v, h_a) Y_i$ which is a weighted mean of the observed values 
with weights given by 
\begin{equation}\label{eq:weig} 
w_{n, i}(v, h) = \frac{ K (h^{-1} (V_i - v))}{
           \sum_{i=1}^n K (h^{-1} (V_i - v)) }.
\end{equation}

\subsubsection{Estimation of $\Gamma_v$.}

For the estimation of $\Gamma_v$ we proceed in an analogous way. Without loss of generality, we assume that $Z$ is centered. First, for $(s, t)\in [0, 1]^2$ fixed,  consider the real valued variables $Y = Z_0(s) Z_0(t)$ and the observations $Y_i = Z_i(s)Z_i(t)$ with $i=1,\ldots,n$. Now redefine the auxiliary quantity $g_v$ using the new definition of $Y$ and $Y_i, \ i = 1, \ldots, n$. Set $g_v(t) = \mathbb{E}^v [Z_0(s) Z_0(t) | V]$ and write the parameter again as the regression of $Y$ against $V$. Then, with a similar reasoning it follows that the estimator of the kernel of $\gamma_v$  at $(s, t)$ is 
  \[ \widehat\gamma_{v, n}(s, t) = 
             \sum_{i=1}^n w_{n, i}(v, h_\gamma) Z_i(s) Z_i(t) ,\]
with weights given by \eqref{eq:weig}. Moreover, on the general case of a not necessarily centered process the estimator of $\Gamma_v$ can be written as
\begin{equation}
  \widehat{\Gamma}_{v, n} = \sum_{i = 1}^n w_{n, i} (v, h_\gamma)
       (Z_i - \widehat a_{v, n}) \otimes (Z_i - \widehat a_{v, n}) .
\end{equation}

\subsubsection{Estimation of the conditional cross covariance operator $\Delta_v$.}

Again, the estimation of the operator is done through the estimation of its kernel, which is in this case the conditional cross covariance function $\delta_v$. We work first with the centered process. Fix $(s, t) \in [0, 1]^2$ and again redefine $Y = Z_0(s)Z_1(t)$ and the observations $Y_i = Z_{i-1}(s)Z_i(t)$ for $i = 2, \ldots, n$. Define $f$ and $g_v$ and their estimators of the same form as \eqref{eq:hatf} and \eqref{eq:hatg} respectively using the bandwidth $h_\delta$ and the new variables $Y$ and $Y_i$. The resulting estimator of $\delta_v(s, t)$ is
 \[ \widehat{\delta}_{v, n} (s, t) = 
          \sum_{i = 2}^n w_{n, i}(v, h_\delta) Z_{i - 1}(s) Z_i (t).\]
We can now plug-in the estimated kernel on the operator which yields the estimator of $\Delta_v$. We write it for the general case of a non centered process as,
   \[   \hat\Delta_{n, v}  = \sum_{i=2}^n w_{n, i}(v, h_\delta) 
                          (Z_{i - 1} - \hat a_n(v)) \otimes (Z_i - \hat a_n(v)) , \]
where the weights are given by equation \eqref{eq:weig}.

\subsubsection*{Remark}
If the denominator on equation \eqref{eq:weig} defining the weights is equal to zero, i.e. $\widehat f_n(v) = 0$, then one usually sets the weights to $w_{n, i}(v, h_a) = n^{-1}$ or $w_{n, i}(v) = 0$ for all $i = 1, \ldots, n$ in order to define the estimator for all $v \in \R^d$. The weights are more important for those segments $Z_i$ with closer value of $V_i$ to the target $v$. The bandwidth plays a key role, tuning the proximity of the scatter of $\R^d$ to $v$ via the scaling of the kernel function. Large values of $h$ lead to weights $w_{n,i}$ that are not negligible for an important number of observations. Conversely, small values result in only few observations having a significant impact on the estimator. This produces the common trade-off between bias and variance of kernel regression estimators.

\subsection{Estimation of $\rho_v$.}

The intrinsic infinite dimension of the space makes difficult the estimation of the operator $\rho_v$. If $H$ is finite-dimensional, the equation (\ref{eq:covrho}) provides a natural way of estimating $\rho_v$. One may plug-in the empirical counterparts of the covariance operators and solve the equation in $\rho_v$. However, when $H$ has infinite dimension, $\Gamma_v$ is not invertible anymore. To well identify $\rho_v$ from \eqref{eq:covrho} the eigenvalues of $\Gamma_v$ need to be strictly positive. An analogous assumption is to ask the kernel of $\Gamma_v$ to be null (see \citet{mas2009linear}). In this case, a linear measurable mapping $\Gamma^{-1}_v$ can be defined as $\Gamma^{-1}_v = \sum_{j\in\N} \lambda_{v, j}^{-1} (e_{v, j} \otimes e_{v, j})$ with domain 
\[  \mathcal{D}_{\Gamma^{-1}_v} = 
     \left\{ z = \sum_{j\in\N} <e_{v, j}, z> e_{v, j} \in H : 
     \sum_{j\in\N}  \left( \frac{<e_{v, j}, z>_H}{\lambda_{v, j}} 
                                 \right)^2 < \infty \right\} , \]
that is a dense subset of $H$. It turns to be an unbounded operator and in consequence continuous nowhere. Hence, there is no hope to obtain any theoretical asymptotic result. However, from \eqref{eq:covrho} we obtain that
  \[ \rho_v^\diamond = \Delta_v \Gamma^{-1}_v, \] 
where $ \rho_v^\diamond$ is the conditional autoregression operator $\rho_v$ restricted to $\mathcal{D}_{\Gamma^{-1}_v}$ as a consequence of $\Gamma_v \Gamma^{-1}_v = I_{\mathcal{D}_{\Gamma^{-1}_v}}$. On the other hand, since the adjoint of a linear operator in $H$ with a dense domain is closed (\textit{closed graph theorem}, see for example \citet[Theorem 5.20]{kato1976perturbation}) and since the range of the adjoint of the cross-covariance operator, $\Delta^{*}_v$, is included in $\mathcal{D}_{\Gamma^{-1}_v}$ we can deduce from \eqref{eq:covrhoADJ} that over $\mathcal{D}_{\Gamma^{-1}_v}$,
             \[\rho^{*}_v = \Gamma^{-1}_v \Delta^*_v.\]
As pointed out by \citet{mas2000estimation} one can use classical results on linear operators to extend by continuity the definition of $\rho_v^\diamond$ to $H$, in order to obtain 
          \[ \rho_v = \text{Ext}(\rho_v^\diamond)
                    = (\Gamma^{-1}_v\Delta^*_v)^* 
                    = (\Delta_v \Gamma^{-1}_v )^{**}.\]
Therefore one may focus on the estimation of $\rho^{*}_v$ because of the theoretical properties are applicable to $\rho_v$ through the composition of $\rho^{*}$ by the adjoint operator.

We can now propose two classes of estimators for $\rho^{*}_v$ (see \citet{mas2000estimation} for analogy with the estimators on the \arh~setting). The first one, the class of \textit{projection estimators}, projects the function space valued observations on an appropriate subspace $H_{v, k_n}$ of finite dimension $k_n = k_{v, n}$. Let $\Pi_{v, k_n}$ be the projector operator over $H_{v, k_n}$. Then one inverts the linear operator defined by the random matrix $\Pi_{v, k_n}\Gamma_{v, n} \Pi_{v, kn}$ and completes with the null operator on the orthogonal subspace. For example, the space $H_{v, k_n}$ may be set equal to the one generated by the first $k_n$ eigenfunctions of $\Gamma_v$. Then, the subspace $H_{v, k_n}$ is estimated by $\widehat{H}_{v, k_n}$, the linear span of the first $k_n$ empirical eigenfunctions. By this way, if $P_{v, k_n}$ is the projection operator on $\widehat{H}_{v, k_n}$, the estimator of $\rho^{*}_v$ can be written as
\begin{equation}\label{eq:rhoproj}
  \widehat\rho_{v, n, k_n}^* = (P_{v, k_n} \widehat\Gamma_{v, n} P_{v, k_n})^{-1} 
                          \widehat\Delta_{v, n}^* P_{v, k_n}.
\end{equation}
The estimation solution by projection over a finite dimensional space is equivalent to approximate $\Gamma^{-1}_v$ by a linear operator with additional regularity $\Gamma^{\dag}_v$ defined as 
\[ \Gamma^\dag_v = 
       \sum_{j = 1}^{k_n} b(\lambda_{v, j}) (e_{v, j} \otimes e_{v, j}),\]
where $(k_n)_n$ is an increasing sequence of integers tending to infinity and $b$ is some smooth function converging point-wise to $x\mapsto 1/x$. Indeed, $\Gamma^\dag_v \to \Gamma^{-1}_v$ when $k_n \to \infty$. The choice of taking $b(x) = 1 /x$ yields, for a finite $k_n$, to set $\Gamma^\dag_v$ equal to a spectral cut of $\Gamma^{-1}_v$. However, this choice is not unique. \citet{mas2000estimation} considers a family of functions $b_{p, \alpha}:\R^+\mapsto\R^+$ with $p\in\N$ such that
    \[ b_{p, \alpha} (x) = \frac{x^p}{(x + \alpha_n )^{p + 1}} ,\]
with $\alpha_n$ a strictly positive sequence that tends to 0 as $n\to+\infty$. With this, the second class of estimators for $\rho^*$, the \textit{resolvent class}, is defined as 
\begin{equation}\label{eq:rhoresolv}
 \widehat\rho_{v, n, p, \alpha}^* = 
         b_{p, \alpha}(\widehat\Gamma_{v, n})  \widehat\Delta_{v, n}^*,
\end{equation}
where we write $b_{p, \alpha} (\widehat\Gamma_{v, n}) = (\widehat\Gamma_{v, n} + \alpha_n I)^{-(p + 1)}$ with $p \geq 0$, $\alpha_n \geq 0$, $n\geq0$. Then, the operator $b_{p, \alpha}(\widehat\Gamma_{v, n})$ can be associated to a regularized approximation of $\Gamma^{-1}_v$ (see \citet{antoniadis2003wavelet} for a discussion on this topic applied to the \arh~estimation). 

Finally, both classes of estimators allow one to predict the future value $Z_{n+1}$ from the observations by first estimating the autocorrelation operator $\rho_v^{*}$ and then applying it to the last available observation $Z_n$.

  \section{Main results}\label{sec:main}

In this section we announce the main theoretical results that justify the choices  made on the estimators  presented in the previous section. 

Neither $(V_k, k \in\Z)$ nor $(Z_k, k \in\Z)$ are assumed to have independent components. We deal with their dependence through a strong mixing hypothesis, that is, we assume each sequence to be asymptotically independent by controlling the decay of the dependence. Many contexts of mixing exist in the literature. In general one relies upon a measure of the decay of a dependence  of two observations as a function of their time gap. We use the 2-$\alpha$-mixing setting, a slightly weaker setting than the $\alpha$-mixing one (see \citet{bosq2007inference}). Let $X = (X_k, k\in\Z)$ be a stationary random process and consider the $\sigma-$algebras $\sigma(X_0)$ and $\sigma(X_k)$ and the 2-$\alpha$-mixing coefficients are defined as
  
  \[ \alpha_X^{(2)} (k)  = 
    \underset{ B\in \sigma(X_0); C \in \sigma(X_k) }{\sup} 
                        |P(B \cap C) - P(B) P(C)| .\]
When $\lim_{k \to \infty} \alpha_X^{(2)} (k) = 0$ we say that $X$ is 2-$\alpha$-mixing. If the mixing coefficients have a geometrical decay, then the corresponding process is called geometrically mixing (GSM). 

\subsection{Convergence of the mean function estimator $a_v$.}

We first prove the pointwise convergence, i.e. for a fixed $t\in[0, 1]$, using the additional Assumptions \ref{ass:punctualconv}. A uniform convergence is obtained by assuming the last two conditions of Assumptions \ref{ass:punctualconv} to hold  uniformly on $[0, 1]$. See \ref{sec:proofs} for proofs together with  explicit constants (depending on $v$) for the convergence rate.

\begin{assumption}\label{ass:punctualconv} Assume that: 
\begin{description}
\item [i.] $V$ admits a probability density function $f$ and for each $s\neq t$, $(V_s, V_t)$ has a density $f_{V_s, V_t}$ such that $\sup_{|s-t|>1} \|G_{s,t}\|_\infty < \infty$ where $G_{s,t} = f_{V_s,V_t} - f\otimes f$.
\item [ii.]Both $(Z_k, k\in\Z)$ and $(V_k, k\in\Z)$ are strong mixing processes with geometrically decaying coefficients $\alpha^{(2)}(k) = \beta_0 e^{-\beta_1 k}$ for some $\beta_0, \beta1 > 0$ and $k \geq 1$.
\item [iii.] $\hilnorm{Z_k} = M_Z < \infty$, $\forall k$.
\item [iv.]The kernel $K$ is a bounded symmetric density satisfying
\begin{enumerate}
 \item $\lim_{v \to \infty} \normRd{v}^d K(v) = 0,$
 \item $\int_{\R^d} \normRd{v}^3 K(v) dv < \infty,$
 \item $\int_{\R^d} |v_i||v_j| K(v) dv < \infty $ for $i,j=1,\ldots,d$. 
\end{enumerate}
\item [v.] The maps $v \mapsto f(v)$ and $v \mapsto g_v(t), t\in [0, 1]$ belongs to $C_d^2 (b)$ the space of twice continuously differentiable functions $z$ defined on $\R^d$ and such that 
\[\left \| \frac{\partial^2 z}{\partial v_i\partial v_j} \right \|_\infty \leq b .\]
\item [vi.] $\E^v[Z_0^2(t) | V]f(v), t\in [0, 1]$ is strictly positive, continuous and bounded at $v\in\R^d$.
\end{description}
\end{assumption}

\begin{prop}\label{teo:a_pointwise}
Under Assumptions \ref{ass:existence} and \ref{ass:punctualconv}, for a bandwidth verifying $h_{a, n} = c_n \left(\frac{\ln n}{n}\right)^{1/(d+4)}$, $c_n\to c>0$, when $n\to\infty$, we have 
\begin{enumerate}
 \item \begin{equation} \label{eq:fconv} 
 \hatf(v) - f(v)    = \mathcal{O}  \left( \left( \frac{\ln n}{n}
       \right)^\frac{2}{4 + d} \right) \qquad \text{a.s.} ,
 \end{equation}
 \item $$ \hata (t) - a_v(t) = \mathcal{O} 
   \left(\left(   \frac{\ln n}{n}
       \right)^\frac{2}{4+d} \right) \qquad \text{a.s.}$$
\end{enumerate}
\end{prop}

Let us comment the assumptions for this result. The density condition 
\ref{ass:punctualconv}(\textbf{i}) may be droped if one uses a more general framework like in \citet{dabo-niang2009kernel} where no density assumption is done and the observations are independent. However, similar results for dependent data are not available yet. The hypothesis concerning the decay of the mixing coefficients allows us to control the variance of the estimators.  We impose some weak conditions on the kernel $K$  that are usual in nonparametric estimation. All symmetric kernels defined over a compact support verify the hypothesis, but also more general ones like the Gaussian kernel. Conditions \textbf{v} and \textbf{vi} are used to control the bias terms of the estimators that is purely analytical.

The convergence rates obtained in Proposition \ref{teo:a_pointwise} are the usual ones. They rapidly degrade with the raise of the dimension of $\R^d$, the space where $V$ lives, as the consequence of the \textit{curse of dimensionality}. In one hand, the first result is well know on the estimation of a multidimensional density functions, even for dependent data. We include it for sake of comprehension. Note that only the observations of $V_1, \ldots, V_n$ are used to estimate $f(v)$ the density of $V$ at $v \in \R^d$. This result is true for each $t\in [0, 1]$. On the other hand, the consistency of $\hata(t)$ is only valid for some fixed value $t \in [0, 1]$. However, we can obtain a version of this result that holds true uniformly on $[0, 1]$ (conditionally on $V$). 

\begin{prop} \label{teo:a_unif}
Under Assumptions \ref{ass:existence},  \ref{ass:punctualconv}\textbf{(i-iv)} and if \ref{ass:punctualconv}\textbf{(v-vi)} hold true for all $t\in[0,1]$, a bandwidth verifying $h_{a, n} = c_n \left(\frac{\ln n}{n}\right)^{1/(d+4)}$, $c_n\to c>0$, when $n\to\infty$, yields
    \[ \hilnorm{ \widehat{a}_n(v,.) - a(v,.) } = \mathcal{O} 
        \left(\left(  \frac{\ln n}{n}\right)^\frac{2}{4+d} \right)
                          \qquad \text{a.s.} \]
\end{prop}

\subsection{Convergence of $\widehat{\Gamma}_{v, n}$ and $\widehat{\Delta}_{v, n}$.}

Similarly to the convergence of the conditional mean function, we first prove the pointwise convergence of $\widehat \gamma_{v, n}(s, t)$ and $\widehat \delta_{v, n} (s, t)$, and then extend the result to the uniform convergence of these kernels over $[0, 1]^2$. Then, the consistency of the operators follows. In addition, we obtain the consistency for the estimators of the spectral elements of $\Gamma_v$.


\begin{prop} \label{teo:gamma}
Under Assumptions \ref{ass:existence} and \ref{ass:punctualconv}, and if 
$\E[\hilnorm{Z}^4 | V] < \infty$ then for a bandwidth verifying $h_{\gamma, n} = c_n \left(\frac{\ln n}{n}\right)^{1/(d+4)}$, $c_n\to c>0$, when $n\to\infty$, we have
  \[ \widehat \gamma_{v, n} - \gamma_v = \mathcal{O} 
   \left(\left(   \frac{\ln n}{n}
       \right)^\frac{2}{4+d} \right) \qquad \text{a.s.} .\]
\end{prop}

Again, the result is valid uniformly for $(t, s) \in [0,1]^2$. Through the equivalence between Hilbert-Schmidt norm and the integral operator norm (on $L_2([0,1]^2)$)
one has, 
\begin{align*}
  \| \widehat\Gamma_{v, n} - \Gamma_v \|_{\mathcal{K}_2}  
   = & \|        \widehat\gamma_{v, n} (.,.)  - \gamma_v (., .) \|_{L^2([0,1]^2)}^2 \\
   = & \Int\Int (\widehat\gamma_{v, n} (s, t) - \gamma_v (s, t)  )^2 ds dt
\end{align*}
and thus the strong consistency of $\widehat \Gamma_{v, n}$ follows.

\begin{prop} \label{teo:unifGamma}
Under Assumptions \ref{ass:existence}, \ref{ass:punctualconv}\textbf{(i-iv)} and if \ref{ass:punctualconv}\textbf{(v-vi)} hold true for all $t\in[0,1]$, and $\E[\hilnorm{Z}^4 | V] < \infty$, then a bandwidth verifying $h_{\gamma, n} = c_n \left(\frac{\ln n}{n}\right)^{1/(d+4)}$, with $c_n\to c>0$, when $n\to\infty$, yields
\[ \| \hat\Gamma_{v, n} - \Gamma_v \|_{\mathcal{K}_2} = \mathcal{O} 
   \left(\left(   \frac{\ln n}{n}
       \right)^\frac{2}{4+d} \right) \qquad \text{a.s.} .\]
\end{prop}

Now, one may use the consistency properties of the empirical eigenvalues $\widehat \lambda_{v, j, n}$ as estimators of the true  ones $\lambda_{v, j}, j\geq 1,$ obtained by \citet{bosq2000linear} in the dependent case. Also a result concerning the convergence of the empirical conditional eigenfunctions $\widehat e_{v, j, n}$ is provided. See \citet{mas2003perturbation} for a general transfer approach of limit theorem properties and modes of convergence from the estimator of a covariance operator to the estimators of its eigenvalues. 

\begin{cor} \label{teo:lambda}
Under the conditions of Proposition \ref{teo:unifGamma}, we have
\begin{enumerate}
 \item \[\sup_{j \geq 1} |\widehat \lambda_{v, j, n} - \lambda_{v, j}| =
            \mathcal{O} \left(\left(
            \frac{\ln n}{n} 
            \right)^\frac{2}{4+d} \right) \qquad \text{a.s.}   \]
 \item \[\hilnorm{\widehat e_{v, j, n}' - e_{v, j}}  =  \xi_{v, j}
             \mathcal{O} \left(\left(
            \frac{\ln n}{n} 
            \right)^\frac{2}{4+d} \right) \qquad \text{a.s.}    \]
\end{enumerate}
where $\widehat e_{v, j, n}' = \innH{\widehat e_{v, j, n}, e_{v, j)}} \widehat e_{v, j, n}$ and $\xi_{v, 1} = 2\sqrt{2} / (\lambda_{v, 1} - \lambda_{v, 2})$, 
$\xi_{v, j} = 2\sqrt{2} / \min (\lambda_{v, j - 1} - \lambda_{v, j}, \lambda_{v, j} - \lambda_{v, j - 1}) $ for $j \geq 2$. 
\end{cor}

Note that the conditional eigenfunctions are estimated up to their sign. This causes problems both in practice and in theory. The estimated object is the eigen-space generated by the associated eigenfunction and not its direction. 

Finally, using similar arguments we obtain the convergence of the conditional cross-covariance operator.

\begin{prop} \label{teo:unifDelta}
Under Assumptions \ref{ass:existence},  \ref{ass:punctualconv}\textbf{(i-iv)} and if \ref{ass:punctualconv}\textbf{(v-vi)} hold true for all $t\in[0,1]$, 
and $\E[\hilnorm{Z}^4 | V] < \infty$, then for a bandwidth verifying $h_n = c_n \left(\frac{\ln n}{n}\right)^{1/(d+4)}$ with $c_n\to c>0$, when $n\to\infty$, yields
\[ \| \widehat\Delta_{v, n} - \Delta_v \|_{\mathcal{K}_2} = \mathcal{O} 
   \left(\left(   \frac{\ln n}{n}
       \right)^\frac{2}{4+d} \right) \qquad \text{a.s.} .\]
\end{prop}

\subsection{Convergence of the predictors}

The two proposed classes of estimators for $\rho^*$ can be use to predict $Z_{n+1}$ by applying them to the last observed function $Z_{n}$. However, since $Z_{n}$ was used on the construction of the estimator and the process has a memory length of 1, a better approach is to study the prediction error on the next element of the sequence. We introduce a final set of assumptions needed to shown the convergence in probability that we denote $\xrightarrow{P}$.

\begin{assumption}\label{ass:pred}\ 
\begin{enumerate}
 \item $\E[\hilnorm{Z}^4 | V] < \infty$.
 \item $\Gamma_v$ is one-to-one.
 \item $\Prob (\liminf \mathcal{E}_n) = 1$, where $\mathcal{E}_n = \lbrace \omega \in \Omega: \dim(\Rg( P^{k_n}_v \widehat{\Gamma}_{v,n} P^{k_n}_v )) = k_n \rbrace$, with $\Rg(A)$ denoting the range of the operator $A$.
 \item $n \lambda_{k_n}^4(v) \to \infty$ and 
 $(1/n)\sum_{j=1}^{k_n} \xi_k (v)/ \lambda_k^2(v) \to0$, as $n\to\infty$.
\end{enumerate}
\end{assumption}

A strong finite fourth conditional moment of $Z$ was used for the definition of $\Gamma_v$ and $\Delta_v$. The second condition in \ref{ass:pred} is necessary to uniquely define the conditional autoregression operator $\rho_v$. The third one is necessary to guarantee that the random operator $P^{k_n}_v \widehat{\Gamma}_{v,n} P^{k_n}_v$ is  almost sure invertible. Controlling the decay of the eigenvalues of the conditional covariance operator is used for the consistency of the projection class operator (see Corollary \ref{teo:lambda} for the definition of $\xi$). Alternatively, one may set $\Lambda_v (k) = \lambda_k(v)$ where $\Lambda_v:\R\to\R$ is a convex function (see \citet{mas2007weak}).

\begin{thm}\label{teo:rhoproj}
If Assumptions \ref{ass:existence}, \ref{ass:punctualconv} hold true $\forall t \in [0,1]$ and \ref{ass:pred}, if $\lambda_k(v) = c_0 c_1^k, c_0>0, c_1\in(0,1)$ and if $k_n = o(\ln n)$ as $n\to\infty$, then
  \[ \hilnorm{ \widehat\rho_{v, n, k_n}^*(Z_{n+1}) - 
     \rho^*(Z_{n+1})}  \xrightarrow{P} 0\]
\end{thm}

\begin{thm}\label{teo:rhoresol}
If Assumptions \ref{ass:existence}, \ref{ass:punctualconv} hold true $\forall t \in [0,1]$ and \ref{ass:pred}(\textbf{i-ii}), and if $b_n\to 0, b_n^{p+2}\sqrt{n}\to\infty$ for some $p\geq0$ as $n\to\infty$, then
  \[ \hilnorm{ \widehat\rho_{v, n, p, \alpha}^*(Z_{n+1}) - 
     \rho^*(Z_{n+1})}  \xrightarrow{P} 0 \]
\end{thm}

  \section{Empirical study}\label{sec:empirical}

We apply the \carh~process model to predict the electricity daily load curve for the french producer EDF (\textit{\'Electricit\'e de France}). Our aim is to introduce the temperature information as an exogenous covariate on a functional prediction model using \carh~processes. The electricity demand is highly sensitive to meteorological conditions. In particular, changes in temperature during winter have a high impact on the French national demand. This relationship is not linear and depends on the hour of the day, the day of the week and the month of the cold season. Moreover, it is unknown in which way the temperature should be coded in order to extract the relevant information for a prediction model. More details on this dataset are given in \citet{antoniadis2013clustering}.

We compare in terms of prediction error, the AutoRegressive Hilbertian model (ARH) and the Conditional AutoRegressive model (CARH). The data we use are the electricity load for the first three months of 2009 (where the load is very sensitive to temperature changes) recorded at a 30 minutes resolution and an estimate of the national temperature computed by EDF recorded each hour. The function-valued process $Z$ is the sequence of daily loads of the national grid. As  the calendar has a very important effect on the electricity demand, we work only with one day-type, namely the weekdays from Mondays to Friday excluding holidays. The covariate $V$ is constructed as an univariate summary of the daily temperature profile. Concretely, we compute the variation coefficient of the temperature records for each day. The total number of observations is 41, where we use the first 33 (approximately 80\%) for calibration of the model and the last 8 to measure the prediction quality of the calibrated model.

\begin{table}[!ht]\centering
\begin{tabular}{lccc}\hline
                  &  ARH  & CARH \\ \hline\hline
Dimension $(k_n)$ &    2  &  5   \\
Estimation error  &  1616 & 929  \\ 
Prediction error  &  1522 & 1265 \\ \hline
\end{tabular}
\caption{Values of the estimated parameters for both ARH and CARH prediction models. The estimated set of bandwidths is $h_a = 1.21\times 10^{-1}, h_\gamma = \times 10^{-4}, h_\delta=3.95\times 10^{-1}$.}
\label{tab:estimatedparameters}
\end{table}

For both models we use projection type estimators (see Equation \eqref{eq:rhoproj}). Using the calibration dataset we estimate the parameter $k_n$, that is the dimension of the projection space for both models. In addition, we estimate the bandwidth parameters for the CARH model. The results of the parameters' estimation is summarised in Table \ref{tab:estimatedparameters}. We also compute the in-sample estimation error as the prediction error obtained using the set of parameters that minimise the root mean square error (RMSE) on the training dataset. The CARH model seems to obtain a better fit on the calibration set since it presents a smaller estimation error.

\begin{figure}[!ht]\centering
\includegraphics[width=0.9\textwidth]{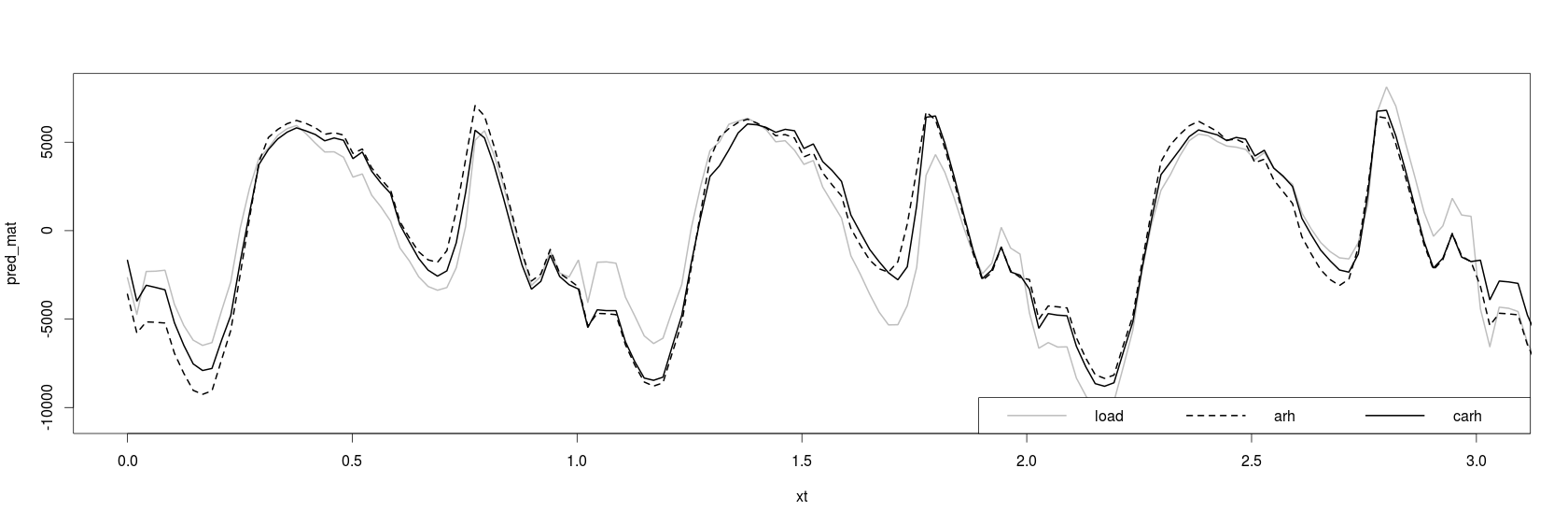}
\caption{Three days of the electricty demand (solid gray) with the  one-day ahead predictions using ARH (dashed) and CARH (solid black) models.}\label{fig:pred}
\end{figure}

In order to estimate the prediction error we use the test dataset. We compute the error as the RMSE. Again the CARH model presents a smaller error than the ARH model. On Figure \ref{fig:pred} we present three days of the electricity demand as well as their predictions using the ARH and CARH models. The effect of the covariate seems to be expressed locally in some parts of the day. Effectively, it corresponds to the daytime demand which seems to be reasonable because the effect of the temperature on the electricity demand is higher during day hours than night hours.


\appendix

  \section{Linear operators in Hilbert spaces}\label{sec:lops}

We recall here some relevant facts about linear operators on Hilbert space (see \citet[Chap. 5]{kato1976perturbation} for details).

We note $H^*$ the topological dual of $H$, i.e. the space of bounded linear functionals on $H$. Thanks to the Riesz representation $H^*$ can be identified with $H$. We note $\mathcal{L}$ the space of bounded linear operators from $H$ to $H$ equipped with the uniform norm 
\[ \supnorm{\rho} = \sup_{\hilnorm{z} \leq 1} \hilnorm{\rho(z)}, \qquad \rho \in \mathcal{L} , z \in H.\]
This space seems to be a too large space, so one usually consider the subspace of compact operators $\mathcal{K}$ that is easier to deal with (see \citet{mas2007weak}). For instance, if the operator $\rho$ is compact then it admits a unique spectral decomposition, i.e. for two bases $(\phi_j)_{j\in\N}$ and $(\psi_j)_{j\in\N}$ and a sequence of numbers $(\lambda_j)_{j\in\N}$ that we can choose to be non-negative (choosing the sign of $\psi_j$) we have
  \[ \rho = \sum_{j\in\N} \lambda_j \psi_j \otimes \phi_j ,\]
where we use the tensor product notation $(u\otimes v)(z) = \innH{u,x} v$ for any elements $z, u, v \in H$. We say that a operator $\rho$ is self-adjoint if $\innH{\rho u, v} = \innH{u, \rho v}$ for all $u, v \in H$. If $\rho$ is symmetric the decomposition becomes $\rho= \sum_{j\in\N} \lambda_j \phi_j \otimes \phi_j$ with eigen-elements $(\lambda_j, \phi_j)_{j\in\N}$. If $\rho$ is not self-adjoint, we call $\rho^*$ its adjoint. Finally we say that $\rho$ is positive-definite if it satisfies $\innH{\rho z, z} \geq 0$ for all $z\in H$. Two subspaces of $\mathcal{K}$ will be of our interest: the space of Hilbert-Schmidt operators $\mathcal{K}_2$ and the space of trace class (or nuclear) operators $\mathcal{K}_1$ defined respectively as
\[ \mathcal{K}_2 = 
      \set{A \in \mathcal{K}: \sum_{j\in\N} \lambda_j^2 < \infty}, 
  \qquad   \mathcal{K}_1 = 
      \set{A \in \mathcal{K}: \sum_{j\in\N} |\lambda_j| < \infty} . \]
The Hilbert-Schmidt operators form a separable Hilbert space with inner product $<\rho, \tau>_{\mathcal{K}_2} = \sum_{j\in\N} <\rho \psi_j, \tau\psi_j>$ with $(\psi_j)_j$ an orthonormal basis and $\rho, \tau \in \mathcal{K}_2$ (the product does not depends on the choice of the basis, see \citet[p. 262]{kato1976perturbation}). The associated norm yields from $\|\rho\|_{\mathcal{K}_2}^2= \sum_{j\in\N} \hilnorm{\rho \psi_j}^2 = \sum_{j\in\N} \lambda_j^2$. On the other hand the space of trace-class operator endowed with the norm $\|.\|_{\mathcal{K}_1}$ defined as $\|\rho\|_{\mathcal{K}_1} = \sum_j |\lambda_j$| is a separable Banach space. Finally, from the continuity of the inclusions $\mathcal{K}_1 \subset \mathcal{K}_2 \subset \mathcal{K} \subset \mathcal{L} $ we have that
 \[ \|.\|_{\mathcal{K}_1} \geq  \|.\|_{\mathcal{K}_2} \geq  \supnorm{.} .\]

  \section{Sketch of proofs.}\label{sec:proofs}

\subsection*{Proof of Theorem \ref{teo:existence}.} 

We mimic the proof of Theorem 1 in \citet{guillas2002doubly}. To prove the existence, Let
\begin{align*}
  \eta_m^{m'} &=
   \E \left\| \sum_{j=m}^{m'} 
       \left(\prod_{p=0}^{j-1} \rho_{V_{n-p}} \right)
                       (\epsilon_{n-j}) \right\|_H^2  \\
   &= \sum_{j=m}^{m'} \E \left\| \left( \prod_{p=0}^{j-1} \rho_{V_{n-p}} \right)
                          (\epsilon_{n-j}) \right\|_H^2 \\
   &\leq \sum_{j=m}^{m'} \E \left[ 
   \left\| \prod_{p=0}^{j-1} \rho_{V_{n-p}}   \right\|_\mathcal{L}^2
   \left\| \epsilon_{n - j}                   \right\|_H^2 \right]  \\
   &\leq \sum_{j=m}^{m'} \E  \left[
   \prod_{p=0}^{j-1} \left\| \rho_{V_{n-p}}  \right\|_\mathcal{L}^2
        \right]
   \underbrace{\E \left\| \epsilon_{n - j} \right\|_H^2}_{ = \sigma^2}
\end{align*}
where we used the independence between $V$ and $\epsilon$ gives
\[ \E \left<
   \left( \prod_{p=0}^{j-1} \rho_{V_{n-p}} \right) (\epsilon_{n-j}), 
   \left( \prod_{p=0}^{j'-1} \rho_{V_{n-p}} \right) (\epsilon_{n-j'})
       \right> = 0, \qquad \text{for } j \neq j' . \]
Finally, we obtain 
\[  \eta_m^{m'} \leq \sigma^2 \E  \left[
     \prod_{p=0}^{j-1} \left\| \rho_{V_{n-p}}  \right\|_\mathcal{L}^2
        \right] \leq \sigma M_\rho^{2j} . \]
We have that the upper bound is the general term of a convergent series. For $m, m'$ tending to infinity, $\eta_m^{m'}$ tend to zero and the Cauchy criterion gives the mean square convergence of the solution.

Now, consider the stationary process $W_n = a +\sum_{j=0}^\infty \left(\prod_{p=0}^{j-1} \rho_{V_{n-p}} \right) (\epsilon_{n-j})$. From the almost surely boundedness of $\rho_{V_n}$ we have that it is indeed a solution of the \carh~process:
\begin{align*}
 (W_n - a ) -& \rho_{V_n} (W_{n - 1} - a)  =  \\
 &=\sum_{j=0}^\infty \left(\prod_{p=0}^{j-1} \rho_{V_{n-p}} \right) \epsilon_{n-j} -
 \sum_{j=0}^\infty \rho_{V_n} \left(\prod_{p=0}^{j-1} \rho_{V_{n-1-p}} \right) \epsilon_{n-1-j} \\
 &= \sum_{j=0}^\infty \left(\prod_{p=0}^{j-1} \rho_{V_{n-p}} \right) \epsilon_{n-j} -
 \sum_{j=0}^\infty \left(\prod_{p=0}^j \rho_{V_{n-p}} \right) \epsilon_{n-1-j} \\
 &= \sum_{j=0}^\infty \left(\prod_{p=0}^{j-1} \rho_{V_{n-p}} \right) \epsilon_{n-j} -
 \sum_{j'=1}^\infty  \left(\prod_{p=0}^{j'-1} \rho_{V_{n-p}} \right) \epsilon_{n-j'} \\
 &= \epsilon_n .
\end{align*}

\subsection*{Proof of Proposition \ref{teo:a_pointwise}}

The proof is based on the classical decomposition in terms of bias and variance of the estimators. The bias term is purely analytical. The variance term is composed by the variance and covariance of the estimator's terms. The dependency of the data is controlled by means of the following exponential inequality (a proof can be founded in \citet[p. 140]{bosq2007inference}).

\begin{lem} Let $W=(W_t)$ be a zero-mean real valued stationary process with $\sup_{1\leq t \leq n} \|W_t\|_\infty = M < \infty$, $(M>0)$. Then for $q\in[1, n/2]$, $\kappa>0$, $\epsilon>0$, $p=n/(2q)$,

\begin{multline}\label{eq:expineq}
 \Prob \left( \left| \sum_{i=1}^n W_i\right| > n\epsilon \right) < 
     \frac{8M}{\epsilon \kappa} (1 + \kappa) 
        \alpha_X \left( \left[ \frac{n}{2q} \right] \right) +  \\
   4 \exp\left(- \frac{n^2\epsilon^2/q}{8(1+\kappa)\sigma(q) + 
               \frac{2M}{3} (1 + \kappa)n^2 q^{-2}\epsilon}\right), 
 \end{multline}
with $\sigma(q)$ an intricate quantity involving the pairwise covariances of $W$. We will only need a bound of $\sigma(q)$ that in the stationary case turns out to be 
\begin{equation}\label{eq:sigmaq}
\sigma(q) < ([p]+2)(\Var(W_0) + 2 \sum_{l=1}^{[p]+1} |\Cov(W_0, W_l)|) . 
\end{equation}
\end{lem}
\textit{Proof of 1.} One has 
\[ \E \hatf(v) - f(v) = \int_{\R^d} K(u)(f(v - h_n u) - f(v))du \]
Using Taylor formula and the symmetry of $K$ one gets
\[ \E \hatf(v) - f(v) = \frac{h_n^2}{2}
      \int_{\R^d} K(u) \left( \sum_{i,j = 1}^d u_i u_j 
       \frac{\partial^2 f}{\partial v_i \partial v_j}(v-\theta h_n u) \right) du \]
where $0<\theta<1$. Finally, Lebesgue dominated convergence theorem gives 
\begin{equation}\label{eq:b_2}
 h_n^{-2} |\E \hatf(v) - f(v)| \to b_2(v) =  
   1/2 \left( \sum_{i,j = 1}^d \frac{\partial^2 f}{\partial v_i \partial v_j}(v) \int_{\R^d} u_i u_j K(u) du \right)
\end{equation}

We use \eqref{eq:expineq} to deal with the variance term $\hatf(v) - \E \hatf(v)$.  Define $W_i = K_h(v - V_i) - \E K_h(v - V_i)$, with $K_h(.) = K(./h)$. Then, $M = 2 h_n^{-d}\|K\|_\infty$. Let us choose $q_n=\frac{n}{2 p_0 \ln n}$ for some $p_0>0$. Which yields on a logarithmic order for $p_n = \frac{1}{p_0 \ln n}$. This choices and the boundeness of $f$ and $G_{s,t}$ entail on \ref{eq:sigmaq},
\begin{align*}
 \sigma(q_n) & < (p_n + 2)\Var(K_h(v-V_1)) + 
                 (p_n + 2)^2 \sup_{|s-t|>1}\|G_{s,t}\|_\infty \\
             & < p_n h_n^{-d} \|K\|_2^2 f(v) (1+o(1)) .
\end{align*}
Now take $\epsilon = \eta \sqrt{\frac{\ln n}{n h_n^d}}$, for some $\eta >0$, then
\begin{multline*}
\Prob \left( n^{-1} \left|\sum_{i=1}^n W_i \right| > \eta \sqrt{\frac{\ln n}{n h_n^d}} \right) < 
 \frac{8 \beta_0 c^{d/2}}{\eta \kappa}(1 + \kappa)\|K\|_\infty   
       \frac{n^{\frac{2+d}{4+d}}-\beta_1 p_0}{(\ln n)^{\frac{2+d}{4+d}}} \\ +
 4\exp\left(-\frac{\eta^2\ln n}{4(1+\kappa)^2\|K\|_2^2 f(v) (1+o(1))}\right)
\end{multline*}

If we take $\eta > 2(1+\kappa)\|K\|_2\sqrt{f(v)}$ and $p_0>2\beta$, then where both terms are $o(n^{-\lambda})$, for some  $\lambda>0$, in which case 
\[ \sum_n \Prob \left\lbrace 
  \left(\frac{n}{\ln n}\right)^{\frac{2}{4+d}} 
  \left|n^{-1} \sum_{i=1}^n W_i \right| > \eta c_n^{-d/2} \right\rbrace
  < \infty. \]
So Borel-Cantelli lemma implies $\limsup_{n\to+\infty} \left(\frac{n}{\ln n}\right)^{\frac{2}{4+d}} |\hatf(v) - \E \hatf(v)| \leq 2c_n^{-d/2} (1+\kappa)\|K\|_2\sqrt{f(v)}$ almost surely for all $\kappa>0$. We have finally
\[   \limsup_{n\to+\infty} \left(\frac{n}{\ln n}\right)^{\frac{2}{4+d}} |\hatf(v) - f_(v)| \leq 2 c_n^{-d/2} \|K\|_2\sqrt{f(v)} + c^2 |b_2(v)|, \]
which gives \eqref{eq:fconv}.

\textit{Proof of 2.} We use the following decomposition, omitting the argument $v$,
\[ \widehat{a}_n - a = \frac{\hatg - a \hatf}{\hatf}.\]
From \eqref{eq:fconv} we have for the denominator that $\hatf \to f(x) \neq 0$ almost surely. We work out the numerator through the following decomposition between variance and bias terms. Let $\psi_n=(n/\ln n)^{2/(4+d)}$, then one has
\[ \psi_n|\hatg - a\hatf| \leq 
     \underbrace{\psi_n|\hatg - a\hatf - \E(\hatg - a\hatf)|}_{:=A_n} +
     \underbrace{\psi_n| \E(\hatg - a\hatf) |}_{:=B_n} .\]

We first study $A_n$ using as before the exponential type inequality \eqref{eq:expineq} with the redefined random variables 
\[W_i =  K_h(v - V_i)(Y_i - a_v) - 
        \E \left( K_h(v - V_i) (Y - a_v) \right)\]
with the precedent choices of $q_n$ and $p_n$. First, one has $|W_i| \leq 2 h_n^{-d} \|K\|_\infty (1+o(1))$. Next, using Bochner lemma (\citet[p. 135]{bosq2007inference}) we obtain 
   \[    h_n^d \Var(W_1) \leq h_n^{-d} \E[K_h^2(v-V_1)(Y_1-a_v)^2] \to
                    f(v) \|K\|_2^2 \Sigma(v) \]
where $\Sigma(v) = (\E^v [Y_0^2 | V] - a_v)$ is the conditional variance parameter. The logarithmic order of $p_n$ and the control on $F$ gives $\sigma^2(q_n) \leq p_n h_n^{-d} f(v) 2 \Sigma(v) \normtwo{K}^2 (1 + o(1))$. As before, taking $p_0>2/\beta_1$ and for a large enough $\eta$, Borel-Cantelli lemma entails
   \[ \limsup_{n\to\infty} A_n \leq  2 c^{-d/2} \sqrt{\Sigma(v)f(v)} 
                  \qquad \text{a.s.} \]

For the bias term we write
  \[    \E(\hatg(v) - a(v)\hatf(v)) = 
       h_n^{-d} \int_{\R^d} K_{h_n} (v - t)(g(t) - f(t)a(v))dt . \]
Then, we use the Taylor formula to expand $g(t) - f(t)a(v)$ and  Assumptions \ref{ass:punctualconv}(\textbf{iii-iv}) to obtain

\[ \psi_n |B_n| \to b_a(v) = \frac{1}{2} \left| \sum_{i,j=1}^d 
 \left\{ \frac{\partial^2 g}{\partial v_i \partial v_j}(v) -
    a(v) \frac{\partial^2 f}{\partial v_i \partial v_j}(v) \right\}
    \int u_i u_j K(u) du
 \right| \]

Finally, putting all the elements together one obtains,
\begin{equation}\label{eq:a_limsup}
\limsup_{n\to\infty} \left(\frac{n}{\ln n}\right)^{\frac{2}{4+d}} 
  |\widehat{a}_n(v) - a(v)| \leq 2 c^{-d/2} \|K\|_2\sqrt{f(v)\Sigma(v)} + c^2 \frac{|b_a(v)|}{f(v)}  
\end{equation}
from with the result is derived.

\subsection*{Proof of Proposition \ref{teo:a_unif}.}

The only terms on equation \ref{eq:a_limsup} that depends on the value fixed for $t$ are the conditional variance parameter $\Sigma$ and the bias $b_a$. With the new hypothesis holding uniformly, for each $v\in\R^d$,  $\Sigma(v, t)$ and $b_a(v, t)$ are bounded uniformly on $[0, 1]$. Then, recalling that
\[ \hilnorm{\widehat{a}_n(v, .) - a(v, .)}^2 = 
   \int_0^1 (\widehat{a}_n(v, t) - a(v, t))^2 dt ,\]
we obtain the derived result.

\subsection*{Proof of Proposition \ref{teo:gamma}.}
The proof follows the same lines that those used to show Proposition \ref{teo:a_pointwise}(2). In particular,
\[ \widehat{r}_n - r = \frac{\hatg - r \hatf}{\hatf}.\]
gives the decomposition between variance and bias terms,
\[ \psi_n|\hatg - r\hatf| \leq 
     \underbrace{\psi_n|\hatg - r\hatf - \E(\hatg - r\hatf)|}_{:=A_n} +
     \underbrace{\psi_n| \E(\hatg - r\hatf) |}_{:=B_n} .\]

Which yields on 
   \[ \limsup_{n\to\infty} A_n \leq  2 c^{-d/2} \sqrt{\Sigma(v)f(v)}
                 \qquad \text{a.s.} \]
where, by the redefinition of $Y$,  $\Sigma(v) = \E^v[(Z_0(s)Z_0(t))^2 | V] - r(v, s, t)$.

Again using Taylor formula to expand $g(t) - f(t)r(v)$ and the precedent Assumptions we obtain
\[ \psi_n |B_n| \to b_r(v) = \frac{1}{2} \left| \sum_{i,j=1}^d 
 \left\{ \frac{\partial^2 g}{\partial v_i \partial v_j}(v) -
    r(v) \frac{\partial^2 f}{\partial v_i \partial v_j}(v) \right\}
    \int u_i u_j K(u) du
 \right|\]

Finally, resembling the terms we get the equivalent of Equation \eqref{eq:a_limsup} with the redefined $\Sigma$ and the bias $b_r$, from with the result is derived.

\subsection*{Proof of Proposition \ref{teo:unifGamma}.}
First, consider the following decomposition
\[\widehat{\Gamma}_{v,n} = \widehat{R}_n(v) 
     - \widetilde{a}_n(v) \otimes \widehat{a}_n(v)    
     - \widehat{a}_n(v)   \otimes \widetilde{a}_n(v)  
     + \widehat{a}_n(v)   \otimes \widehat{a}_n(v)     ,\]
where $\widehat{R}_n(v) = \sum_{i=1}^n w_{n,i}(v, h_\gamma) Z_i \otimes Z_i$ is the empirical counterpart of the second order moment operator  $R(v) = \E^v[Z_0 \otimes Z_0 |V]$, and $\widetilde{a}_n(v) = \sum_{i=1}^n w_{n,i}(v, h_\gamma) Z_i$.  
Second, we obtain that
\[\Gamma_v - \widehat{\Gamma}_{v,n} = 
     R(v) - \widehat{R}_n(v) -  a(v) \otimes a(v)
     + \widetilde{a}_n(v) \otimes \widehat{a}_n(v)    
     + \widehat{a}_n(v)   \otimes \widetilde{a}_n(v)  
     - \widehat{a}_n(v)   \otimes \widehat{a}_n(v)     .\]
Hence, we can control the estimation error regrouping the terms of the above decomposition (we drop the argument $v$),
\begin{equation}\label{eq:decompGamma}
 \hsnorm{ \Gamma_v - \widehat{\Gamma}_{v,n} } 
    =  \hsnorm{R - \widehat{R}_n}    
    +  \hsnorm{ \widetilde{a}_n \otimes \widehat{a}_n    
                   -  a \otimes a }
    +  \hsnorm{ \widehat{a}_n  \otimes
             (\widetilde{a}_n - \widehat{a}_n) } .
\end{equation}
From Propositions \ref{teo:gamma} and \ref{teo:a_unif} it follows that 
\[ \hsnorm{R - \widehat{R}_n} =   
      \mathcal{O} \left(
       \left(\frac{n}{\ln n}\right)^{\frac{2}{4+d}}
                 \right) \qquad \text{a.s.} \]
The second term of the left hand side of equation \eqref{eq:decompGamma} is equal to
\[\hsnorm{ \widetilde{a}_n \otimes (\widehat{a}_n - a) +
           (\widetilde{a}_n - a) \otimes a } \leq 
  \hilnorm{ \widetilde{a}_n } \hsnorm{ \widehat{a}_n - a } +
    \hsnorm{ \widetilde{a}_n - a} \hilnorm{a} .
\]
Since both $\hilnorm{a}$ and $\hilnorm{ \widetilde{a}_n }$ are bounded and using Proposition \ref{teo:a_unif} successively for  $\widetilde{a}_n$ and $\widehat{a}_n$ with their respective sequences of bandwidths $h_{\gamma,n}$ and $h_{a,n}$, we obtain that 
  \[\hsnorm{ \widetilde{a}_n \otimes \widehat{a}_n    
                    -  a \otimes a } =
  \mathcal{O} \left(
       \left(\frac{n}{\ln n}\right)^{\frac{2}{4+d}}
                 \right) \qquad \text{a.s.} \]
With a similar reasoning, the same kind of result is obtained for the third term in \eqref{eq:decompGamma}. Putting the result for the three terms together conclude the proof.

\subsection*{Proof of Corollary \ref{teo:lambda}.}
First item is a direct consequence of the following property on eigenvalues of compact linear operators \citet[p. 104]{bosq2000linear},
\[ \sup_{j\geq1} |\lambda_j(v) - \hat\lambda_{j,n}(v) | \leq
   \supnorm{\Gamma_v - \widehat{\Gamma}_{v,n}} , \]
and the asymptotic result obtained for $\hsnorm{\Gamma_v - \widehat{\Gamma}_{v,n}}.$

For the second item, Bosq (2000, Lemma 4.3) shows that, for each 
$j\geq 1$,
\[ \hilnorm{e_j(v) - e'_{j,n}(v)}  \leq
    \xi_j \supnorm{\Gamma_v - \widehat{\Gamma}_{v,n}} . \]
Again, the rates of convergence follows from Proposition \ref{teo:unifGamma}.

\subsection*{Proof of Proposition \ref{teo:unifDelta}.}

The proof follows the same guidelines that those of Proposition \ref{teo:unifGamma}, replacing $\hat R(v)$ and $R(v)$ by $\hat R_1(v) = \sum_{i=1}^{n-1} w_{n,i}(v, h_\gamma) Z_i(s)Z_{i+1}(t)$ and $R_1(v)=\E^v[Z_0(s)Z_1(t)|V]$ respectively.
Then, a decomposition like \ref{eq:decompGamma} and the same kind of observations done for that proof entails the result.

\subsection*{Proof of Theorem \ref{teo:rhoproj}.}
The proof follows along the same lines of Proposition 4.6 in \citet{bosq1991modelization} by using Propositions \ref{teo:a_unif}, \ref{teo:unifGamma}, \ref{teo:unifDelta} and Corollary \ref{teo:lambda}.

\subsection*{Proof of Theorem \ref{teo:rhoresol}.}
The proof follows along the same lines of Proposition 3 in \cite[Chapter 3]{mas2000estimation} by using Propositions \ref{teo:a_unif}, \ref{teo:unifGamma} and \ref{teo:unifDelta}.



\bibliographystyle{model1-num-names}
\bibliography{cugliari-jma}







\end{document}